\documentclass[review,3p, 12pt]{elsarticle}

\usepackage{amsmath,amsmath,mathrsfs}
\usepackage{color,fancyhdr,epsf,psfrag,lineno,hyperref,txfonts,subfigure}
\modulolinenumbers[5]
\graphicspath{ {figs/} }
\usepackage{bm,ulem}

\definecolor{R1}{rgb}{0.0,0.0,0.0} 
\definecolor{R3}{rgb}{0.0,0.0,0.0} 
\definecolor{R2}{rgb}{0.0,0,0.0} 

\journal{ }

\biboptions{sort&compress}

\begin{document}

\begin{frontmatter}

\title{Numerical investigation of azimuthal thermoacoustic instability in a gas turbine model combustor}

\author[a,b]{Zhi X. Chen\corref{cor1}}
	\cortext[cor1]{Corresponding author.}
	\ead{chenzhi@pku.edu.cn}
\author[c]{Nedunchezhian Swaminathan}
\author[d]{Marek Mazur}
\author[e]{Nicholas A. Worth}
\author[f]{Guangyu Zhang}
\author[f]{Lei Li}

\address[a]{State Key Laboratory of Turbulence and Complex Systems, Aeronautics and Astronautics, College of Engineering, Peking University, Beijing 100871, P.R. China}
\address[b]{AI for Science Institute (AISI), Beijing, 100080, P.R. China}
\address[c]{Department of Engineering, University of Cambridge, Trumpington Street, Cambridge CB2 1PZ, UK}
\address[d]{CORIA-UMR 6614 Normandie Universit\'e, CNRS-Universit\'e et INSA de Rouen, Campus Universitaire du Madrillet, Saint Etienne du Rouvray, France}
\address[e]{Department of Energy and Process Engineering, Norwegian University of Science and Technology, N-7491 Trondheim, Norway}
\address[f]{Research Institute of Aero-Engine, Beihang University, Beijing 100191, P.R. China}

\begin{abstract}
	Self-excited spinning mode azimuthal instability in an annular combustor with non-swirling flow is investigated using large eddy simulation (LES). Compressible Navier-Stoke equations are solved with a flamelet combustion model to describe the subgrid chemistry$-$turbulence interactions. Two flamelet models, with and without heat loss effects, are compared to elucidate the non-adiabatic wall effects on the thermoacoustic instability. The azimuthal modes are captured well by both models with only marginal differences in the computed frequencies and amplitudes. By comparing with the experimental measurements, the frequencies given by the LES are approximately 10\% higher and the amplitudes are well predicted. Further analysis of the experimental and LES data shows a similar dominant anti-clockwise spinning mode, under which a good agreement is observed for the phase-averaged heat release rate fluctuations. Dynamic mode decomposition (DMD) is applied to shed more light on this spinning mode. The LES and experimental DMD modes reconstructed for their azimuthal mode frequencies agree very well for the heat release fluctuations. The DMD mode structure for the acoustic pressure from the LES shows a considerable non-zero profile at the combustor outlet, which could be essential for azimuthal modes to establish in this annular combustor. Finally, a low-order modelling study was conducted using an acoustic network combined with the flame transfer function extracted from LES. The results show that the dominant mode is associated with the plenum showing a first longitudinal and azimuthal mixed mode structure. By tuning the plenum length to match the effective volume, the predicted frequency becomes very close to the measured value. 
\end{abstract}


\end{frontmatter}


\section{Introduction}\label{sec:introduction}

Azimuthal thermoacoustic instabilities remain an unresolved issue for gas turbine (GT) industries~\cite{LieuwenY06,Poinsot17}. 
This instability arises when the unsteady combustion of multiple circumferentially aligned flames couples with the acoustic resonant modes of the annular chamber, which often dominates over other modes exhibiting higher oscillation frequencies and amplitudes~\cite{OConnorPECS15}.
As pressure waves propagate not only in the longitudinal direction but also in the transverse direction along the circumference, azimuthal modes involve more complex dynamics such as flame/flame interaction~\cite{DawsonCNF14}, modal dynamics~\cite{WorthCNF13}, azimuthal/longitudinal mode coupling (e.g., the \textit{slanted mode})~\cite{BourgouinPCI15}, etc. 
These complexities make azimuthal modes very difficult to understand, predict and control. 

To simulate and predict thermoacoustic instabilities, a range of approaches have been developed over the past several decades, from hand-calculated original analytical~\cite{CroccoJARS51,CroccoJARS52} models, then desktop running low-order network models, to large eddy simulation (LES) now performed on supercomputers, accompanied with a vast increase of computational cost~\cite{Poinsot17}.
Due to the complex burner geometry and boundary conditions, analytical and low-order network models still play essential roles in studying azimuthal modes, see Refs.~\cite{BauerheimPF16,DowlingS03} for extensive reviews on these models. 
To reduce the computational burden, hybrid approaches combining \textit{incompressible} LES and acoustic solvers have also been developed~\cite{HanCNF15b,BauerheimPCI15}, where the connection between solvers, i.e., the flame response to acoustic perturbations is considered through a flame transfer or describing function (FTF or FDF).
The advantage of this approach is that the flame calculation, e.g. reactive flow LES, is decoupled from the system acoustics calculations so that it can still be performed using the computationally much cheaper low-Mach formulations.  
The major difficulty, however, is how to apply appropriate forcing to the flame when the nozzle inlet involves complex geometry with multiple swirlers, or the perturbations need to be applied in transverse directions for azimuthal instabilities. 
With the rapid advances in high-performance parallel computing over recent years, compressible LES for full annular lab-scale burners has now become computationally feasible~\cite{WolfSGMP12,ZettervallPCI19}, offering a promising alternative to investigate azimuthal instabilities and the complex underlying physical mechanisms for modal dynamics and structures. 

Due to the high cost of running annular test-rigs, few laboratory annular combustor experiments are available (e.g., \cite{WorthCNF13,BourgouinASME13}). 
It has been shown in these studies that the swirl-generated azimuthal mean flow can introduce complex time- and operating condition-dependent switching among spinning, standing and mixed modes. 
To better understand the modal behaviour of spinning mode without azimuthal bulk flow, Mazur \textit{et.~al}~\cite{MazurPCI19} removed the swirlers from the 12-burner annular rig studied in~\cite{WorthCNF13,DawsonCNF14} and showed well-characterised spinning modes occurring for high amplitude instabilities over  a wide range of operating conditions. 
Therefore, these experiments provide an ideal test case for LES of azimuthal instabilities.  
The main objective of this study is to simulate the non-swirling annular premixed combustor of Mazur \textit{et.~al}~\cite{MazurPCI19} and explore the LES capabilities of capturing spinning modes observed in the experiments. 
Moreover, previous LES studies have been performed over relatively short run times (tens of ms, much shorter than experiments). The present study also investigates the effect of running for extended times in order to evaluate the time varying mode development.

Amongst the available subgrid combustion models, flamelet approaches are preferable for complex geometries because of their robustness and low computational cost. 
However, compared to models that compute chemistry explicitly (not necessarily refer to computing chemistry using resolved quantities known as \textit{no model} or \textit{quasi-laminar chemistry}, but also in thickened flame and partially stirred reactor models, where local temperature and pressure effects are inherently considered in reaction rate closure), tabulation models require additional non-trivial modelling treatment to account for heat loss effects, which were shown to have a strong impact on the FTF/FDF~\cite{BauerheimPCI15}. 
More recent studies by Kraus~\textit{et~al.}~\cite{KrausCNF18} and Agostinelli~\textit{et~al.}~\cite{AgostinelliCNF21} employed advanced conjugated heat transfer method to account for the two-way coupling between the flame LES and burner walls. They showed that the heat loss effects are important for the swirl burners considered and need to be taken into account in the LES in order to correctly capture the thermoacoustic instability. 
Thus, a non-adiabatic flamelet approach is adopted from Massey~\textit{et~al.}~\cite{MasseyFTC21} for this work and the heat loss effects on the instability simulation are investigated.

This paper is organised as follows. The target experimental configuration is described in Section~\ref{sec:experiments}, followed by the LES modelling framework and numerical setup in Section~\ref{sec:modelling}. The LES results are discussed in Section~\ref{sec:results} and the low-order modelling is presented in Section~\ref{sec:low_order}. The conclusions are summarised in Section~\ref{sec:conclusions}.


\section{Experimental test case}   \label{sec:experiments}

The target annular combustor was first experimentally studied at Cambridge by Worth and Dawson~\cite{WorthCNF13,DawsonCNF14} to characterise and understand azimuthal modes in laboratory-scale swirled burners at atmospheric pressure. By varying the number of burners (i.e., flame-to-flame spacing) and operating flow conditions, various azimuthal modal dynamics were observed involving complex coupling and interplay among different physical processes, most of which still remain to be understood. 
To simplify the system and thus better understand the individual processes, this combustor was recently modified by Mazur~\textit{et.~al}~\cite{MazurPCI19} at NTNU by removing the swirlers in the 12 flames configuration (largest flame spacing). As such, azimuthal modes could be established in the absence of azimuthal mean flow and flame$-$flame interactions, in a geometrically simpler flow.

\begin{figure*}[!th]
\centering\centering\ifx\mycmd\undefined
\includegraphics[width=\textwidth]{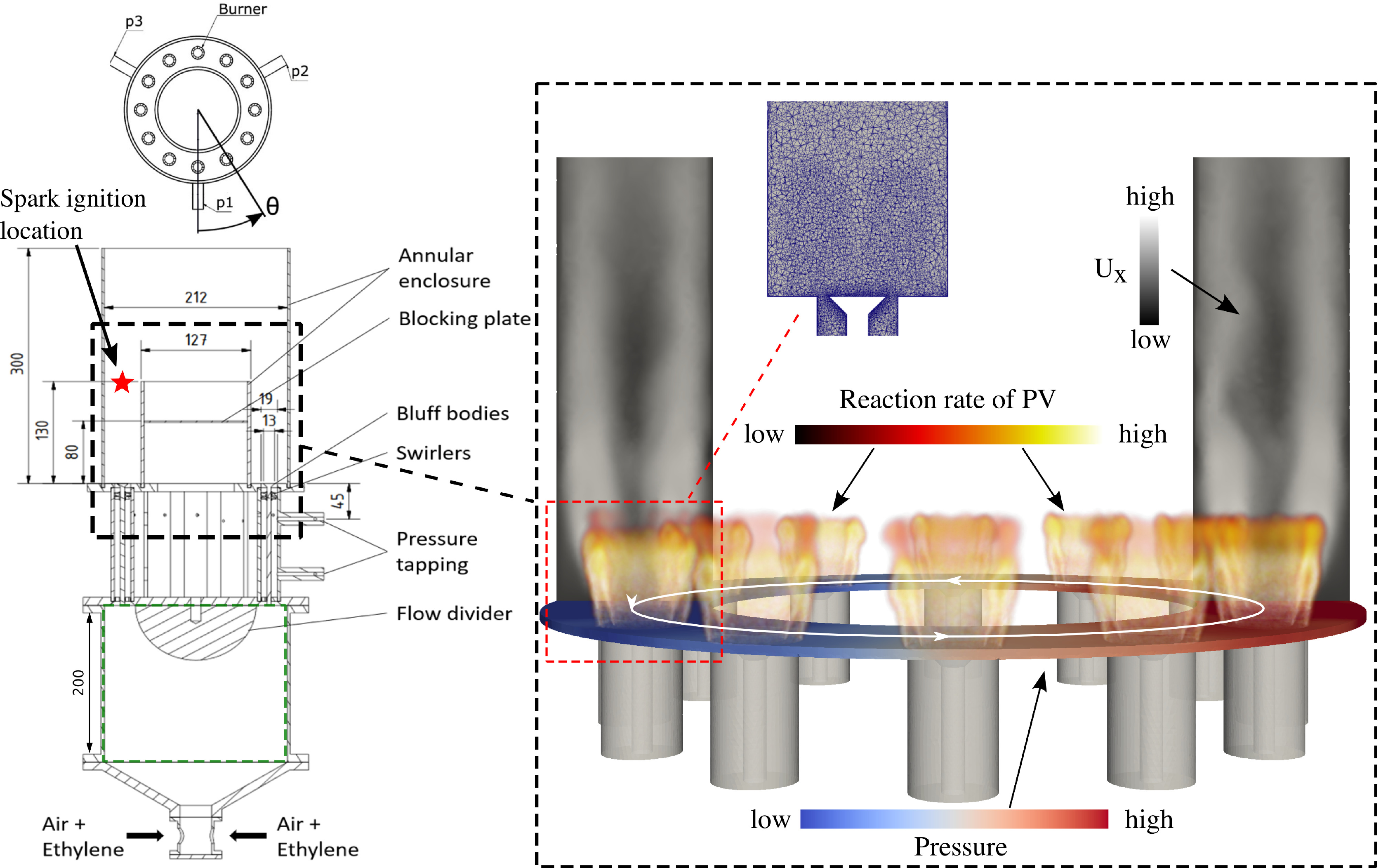}
\fi
\caption{Mid-plane cut view of the experimental configuration~\cite{MazurPCI19} (left) and typical LES snapshot undergoing ACW spinning mode (indicated by white arrows). Zoomed view of the mesh near bluff-body is also shown. } 
\label{fig:exp_setup}
\end{figure*} 

The experimental geometry has been described in detail in the previous work~\cite{WorthCNF13}, and is only summarised briefly here together with important changes. A schematic of the modified burner is shown in Fig.~\ref{fig:exp_setup} with all the dimensions in mm. 
A perfectly premixed ethylene-air mixture is passed through a cylindrical plenum chamber, before being divided around a central bluff body into 12 injector tubes. Each injector tube holds a central rod, to which a conical bluff is mounted  with a diameter of 13~mm. 
In the present work, a bulk flow velocity of 24.5~m/s and a equivalence ratio of 0.95 are employed. Further to Ref.~\cite{MazurPCI19}, the geometry was slightly modified in order to more closely replicate the simulation boundary conditions. In particular, all flow conditioning elements were removed from the plenum and a high level of uniformity was demonstrated suggesting that the conditioning provided by the plenum is sufficient.
An additional plate was mounted inside the inner annular wall cylinder (marked as `blocking plate' in Fig.~\ref{fig:exp_setup}) in order to prevent the entrainment of ambient air through the central opening during the experiment. 
Measurements~\cite{MazurPCI19} were performed using dynamic pressure sensors ($f_{s}=51.2$~kHz, sampled over $t_s=10$~s) at 3 equi-spaced angular locations, as shown in Fig.~\ref{fig:exp_setup}. Pressure ports were flush-mounted in the injector tubes, at a distance $45$~mm upstream of the dump plane. Heat release rate imaging is performed using a high speed intensified camera ($f_{c}=10$~kHz, sampled over $t_c=2$~s, spatial resolution 3.2 pixels/mm).

The computational domain includes the full geometry of the combustor including the upstream plenum, 12 premixers (feeding tubes), combustion chamber and the downstream atmospheric buffering domain. Note that the conical volume at the bottom of the plenum was omitted and only the cylindrical part (marked by green in Fig.~\ref{fig:exp_setup}) is considered to simplify the inflow conditions. However, this simplification has an considerable impact on the predicted mode frequencies, which will be discussed in detailed in Section~\ref{sec:low_order}.
A typical snapshot of LES undergoing an anti-clockwise (ACW) spinning mode is also shown in Fig.~\ref{fig:exp_setup}. Chamber pressure contours are plotted with the arrows indicating the direction of the travelling azimuthal wave. The oscillating flame surfaces are marked using volume rendering for the reaction rate of progress variable. 
The axial velocity contours in the bluff-body mid-plane are shown for two opposite burners.  
From one flame to another along the annulus, the flame length varies in response to the acoustic perturbations resulting in azimuthal heat release fluctuations, which give rise to further acoustic perturbations in the feedback loop. 
To capture this phenomenon, an appropriate LES framework considering compressibility effects is required along with correct numerical boundary conditions for the acoustics.
In addition, a robust and efficient SGS combustion model is also essential and these are described next.


\section{LES framework and numerical setup}   \label{sec:modelling}


The LES modelling approach used for this study involves the flamelet chemistry of 1D freely propagating laminar premixed flames and a presumed probability density function (PDF) to account for the subgrid scale (SGS) turbulence$-$chemistry interactions. This approach has been previously tested and validated extensively for a broad range of flames and configurations. In particular, it was shown in~\cite{ChenPCI19,ChenCNF19} that the unstable flame dynamics and a longitudinal (Helmholtz) mode thermoacoustic instability in a dual swirl GT burner were well captured.
Thus, it is of interest here to further assess this approach for azimuthal modes in annular combustors.
The modelling framework is briefly described below and see details in \cite{ChenCNF19} and Refs. therein. 

Compressible 3D Navier-Stokes equations are solved for filtered mass and momentum in a pressure-based form for the low-Mach number flows considered in this study: 
\begin{flalign} \label{eq:mass}
\frac{\partial \, \overline\rho}{\partial t} + \nabla \cdot \left(\, \overline \rho\,  \widetilde{\bm U}\, \right) &  = 0\,, &  
\end{flalign} 
\begin{flalign} \label{eq:momentum}
\frac{\partial \, \overline\rho\, \widetilde{\bm U}}{\partial t} + \nabla \cdot \left(\, \overline \rho \, \widetilde{\bm U}\,\widetilde{\bm U}\, \right) & = - \nabla\, \overline p ~+~  \nabla \cdot {\bm \tau}_{\rm eff}\,, &  
\end{flalign} 
where the modified filtered pressure $\overline p$ is computed by solving a Poisson equation~\footnote{It is worth noting that here the formulation used is different from the so-called \textit{variable-density low Mach solver}, in which pressure and density are solved in a decoupled manner and hence density variation is only a result of varying temperature. The implementation of $\rho$-PIMPLE algorithm in OpenFOAM-v7 is employed and it has been shown in previous studies~\cite{ChenCNF19,SuASME15} that this algorithm is able to capture aero- and thermo-acoustic phenomena accurately.}. The filtered mixture density, $\overline \rho$, is obtained using the ideal-gas equation of state considering pressure variations.
The effective shear stress tensor is given by 
${\bm \tau}_{\rm eff} = 2\overline\rho\left(\,\widetilde \nu + \nu_{t}\right)\left[\widetilde {\cal {\bm S}} - (\nabla \cdot \widetilde{\bm U})\,{\textbf I}/3\right]$, where $\widetilde {\cal {\bm S}}$ is the filtered strain rate, $\widetilde \nu$ and $\nu_{t}$ are the filtered molecular and subgrid eddy viscosities respectively.
The latter is computed using the Smagorinsky model~\cite{Smagorinsky63}.
To describe SGS turbulent premixed combustion with heat losses, the non-adiabatic flamelet thermochemical states are pre-tabulated using three control variables: filtered reaction progress variable, $\widetilde c$, its SGS variance, $\widetilde{c''^2}$, and total enthalpy, $\widetilde h$ (sensible$+$formation), which are solved using their transport equations: 
\begin{flalign} \label{eq:transp_c}
\frac{\partial \, \overline\rho\, \widetilde{c}}{\partial t} + \nabla \cdot \left( \,\overline \rho \, \widetilde{\bm U}\;\widetilde{c}\, \right)& =  \nabla\cdot \left(\, \overline\rho {\mathscr D}_{\rm eff} \nabla\,\widetilde{c}\;\right)\,+\,\overline{\dot\omega_{c}}\;,  & 
\end{flalign}
\begin{flalign} \label{eq:transp_cvar}
\frac{\partial \, \overline\rho\, \widetilde{c''^2}}{\partial t} + \nabla \cdot \left(\, \overline \rho \, \widetilde{\bm U}\,\widetilde{c''^2}\, \right)  =\; &\nabla\cdot \left(\, \overline\rho {\mathscr D}_{\rm eff} \nabla\,\widetilde{c''^2}\;\right)\,- 2\,\overline{\rho}\,\widetilde{\chi}_{{\rm sgs}}\, \nonumber \\ &
\hspace{-15mm} +2\,\overline\rho\frac{\nu_t}{{\rm Sc}_t}|\nabla\, \widetilde c\,|^2 +2\left(\,\overline{c\,{\dot{\omega}_{c}}}-\widetilde{c}\,\overline{\dot\omega_{c}}\,\right)\,, \hspace{-10mm} & 
\end{flalign}
\begin{flalign} \label{eq:enthalpy}
\frac{\partial \, \overline\rho\, \widetilde{h}}{\partial t} + \nabla \cdot \left(\, \overline \rho \, \widetilde{\bm U}\,\widetilde{h}\, \right) & = \nabla\cdot \left[ \,\overline \rho \left(\frac{\widetilde{\nu}}{\rm Pr} + \frac{\nu_t}{{\rm Pr}_t} \right)\nabla\widetilde{h}\,\right] + {\frac{D \overline p}{Dt}} \,, \hspace{-5mm} & 
\end{flalign}
where the laminar and turbulent Prandtl numbers, ${\rm Pr}$ and ${\rm Pr}_t$, are both taken to be 0.7.
The effective diffusivity ${\mathscr D}_{\rm eff}$ is modelled using $(\,\widetilde D + \nu_t/{\rm Sc}_t)$ with a turbulent Schmidt number ${\rm Sc}_t = 0.4$~\cite{ChenRS16b,PitschS00}, 
and $\widetilde D$ is the filtered molecular diffusivity.  
The progress variable is defined using the sum of CO and CO$_2$ mass fractions and it is normalised to vary from 0 to 1 between the fresh and burnt gases.
The unclosed reaction rate and dissipation rate terms are modelled through a revised flamelet formulation as detailed in Refs.~\cite{ChenCNF19,LangellaJPP18}

\begin{figure}[!h]
\small
\centering\ifx\mycmd\undefined
\includegraphics[width=0.7\columnwidth]{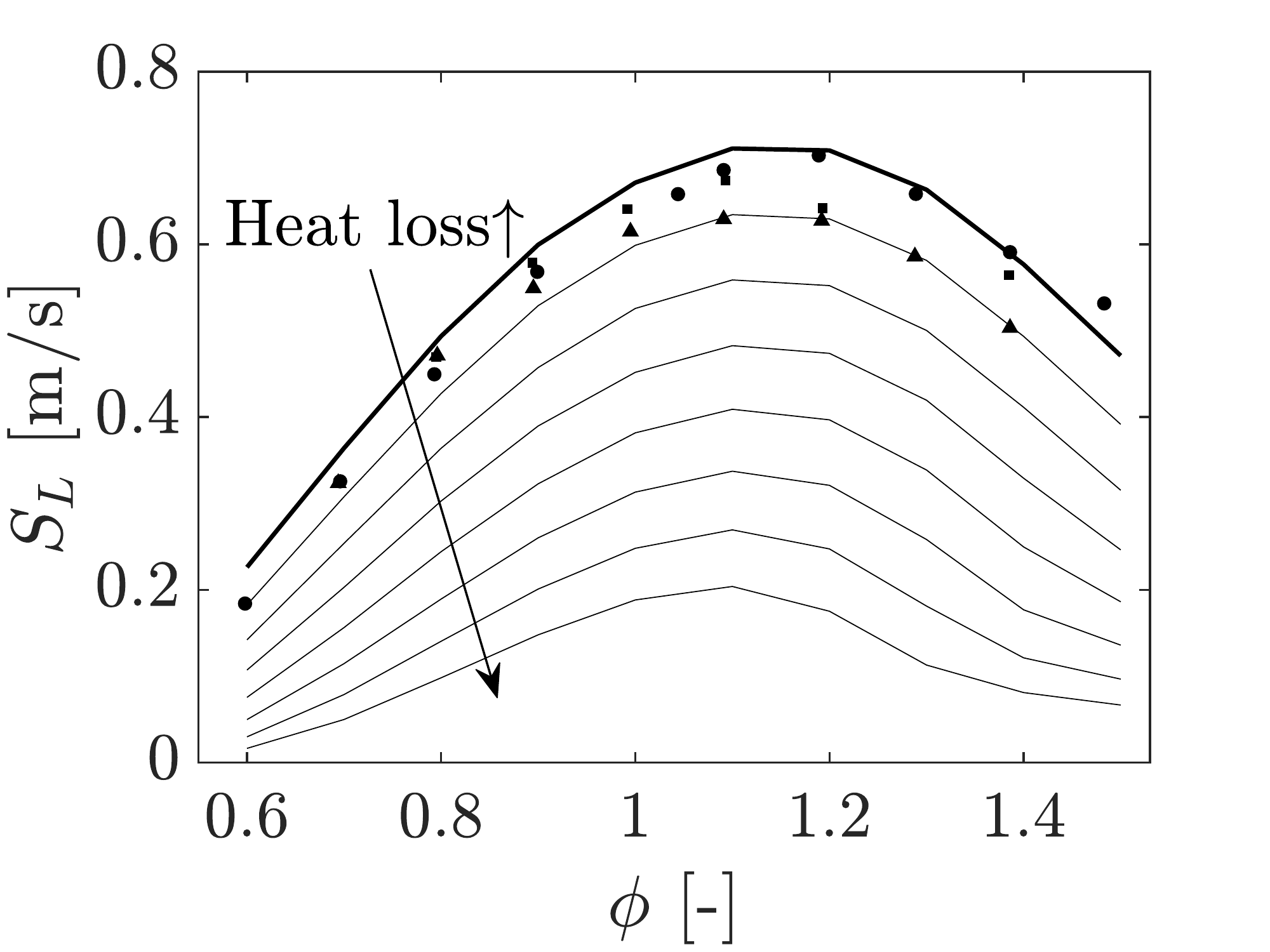}
\fi
\caption{Laminar flame speed versus equivalence ratio for ethylene/air mixture at 300~K and 1~atm. Symbols: measurements~\cite{XuEnergy12}; Thick line: adiabatic simulation; Thin lines: simulations with heat losses. } 
\label{fig:SLvsPhi}
\end{figure}

\begin{figure}[!h]
\small
\centering\ifx\mycmd\undefined
\includegraphics[width=0.8\columnwidth]{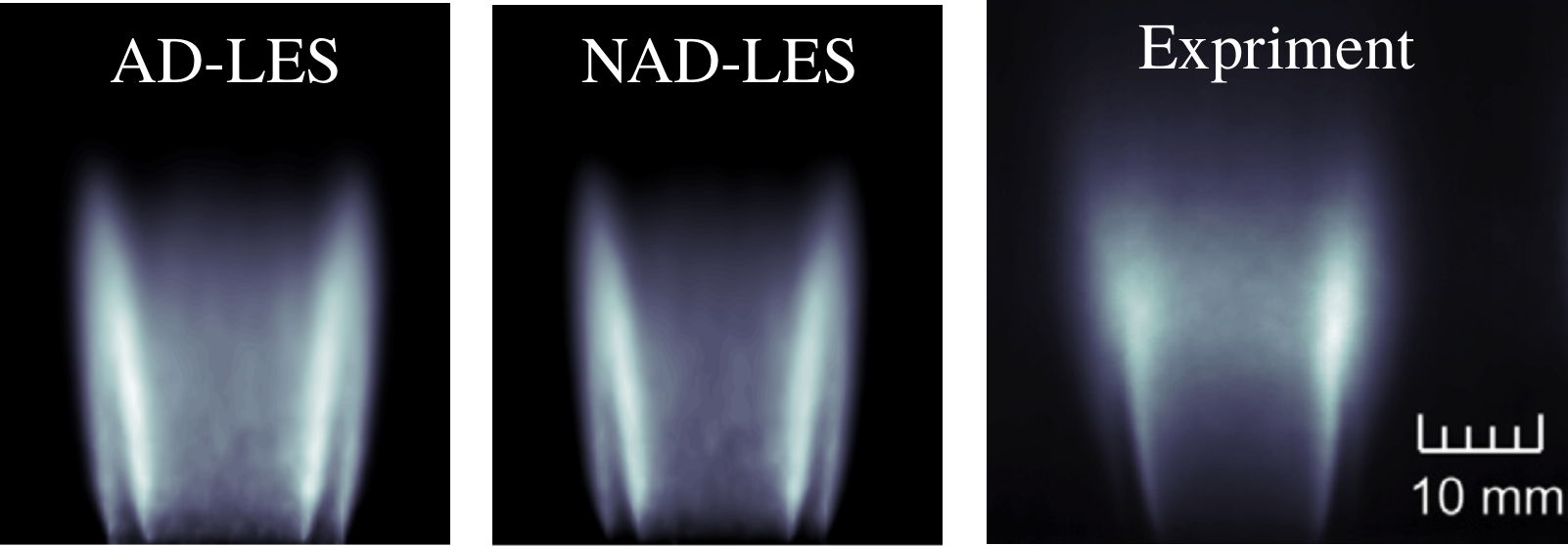}
\fi
\caption{Mean shape of single flame: line-of-sight integrated LES heat release rate using adiabatic (left) and non-adiabatic (middle) flamelets, compared with the experimental~\cite{MazurPCI19} OH$^*$ image (right).  } 
\label{fig:meanFlame}
\end{figure}

Laminar 1D premixed flame are calculated using Cantera for a range of heat loss levels. The full USC mechanism (C1$-$C3)~\cite{WangUSC98} is used for ethylene/air combustion at 300~K and 1~atm.  The simulation results are plotted against the experimental measurements~\cite{XuEnergy12} in Fig.~\ref{fig:SLvsPhi}. 
Heat loss effects are implemented in a similar fashion to that used in Ref.~\cite{MasseyFTC21} by scaling down the heat release rate in the energy equation. 
As can be seen in the figure, the computed adiabatic flame speeds agree well with the measurements and as the heat loss increases, the flame speed drops smoothly. Figure~\ref{fig:meanFlame} compares the measured and computed mean flame shapes using time-averaged OH$^*$ chemiluminescence from the experiments and heat release rate from LES with non-adiabatic (NAD) and adiabatic (AD) flamelet models. Note that since the OH$^*$ image is a line-of-sight visualisation, the LES results are integrated in the transverse direction. 
It is seen that the flame shape and length are well captured in the LES. In comparison with the AD model, the NAD case shows weaker burning outer branches with extinguished flame roots near the dump plane, which is similar to the experimental image. 
Bauerheim~\textit{et.~al}~\cite{BauerheimPCI15} argued that the lifted flame base is more prone to acoustics-induced oscillations, which may have a impact on the flame response for azimuthal modes. This is further discussed later in Section~\ref{subsec:comparison}.

The LES modelling framework described above is implemented using the OpenFOAM 2.3.0 libraries.
The computational domain includes the upstream tubes and plenum, and a fixed mass flow rate condition ($u'=0$) is employed for the inlet boundary. To mimic the open downstream boundary, a large hemispherical domain is added and the far-field boundary is specified as a pressure outlet ($p'=0$). 
A fully unstructured grid with about 18 million ($\sim$1.5M per burner) tetrahedral cells is used and a zoom-in view of the mesh near the bluff-body is shown in Fig.~\ref{fig:exp_setup}.
Mesh sensitivity was investigated for a single sector using a finer mesh and similar results were found for the mean fields and also the FTF (see Appendix A). 
Isothermal condition of 1000~K~\cite{ZettervallPCI19,BauerheimPCI15} is given for the combustion chamber and bluff-body top walls and adiabatic condition is specified for all other walls.
The Time-step is determined such that CFL $<$ 0.5 for the entire domain and the typical size is between $2-5\times10^{-7}$~s.
A fully developed cold flow is first simulated for about 200~ms before ignition and then there is a transient period for azimuthal modes to settle and subsequently establish limit-cycle oscillations.
The length of this transient period ranges between tens and hundreds of ms, highly depending on the ignition method used and this is discussed next.


\section{Results and discussion}\label{sec:results}

\subsection{Onset of computed azimuthal modes: effect of the ignition method} 

For LES of multi-sector combustors, ignition is commonly conducted by duplicating the reactive fields of a separate single-sector simulation to reduce the cost~\cite{ZettervallPCI19}. 
This differs from the experiments~\cite{MazurPCI19} where an ignition source was used to light the burner from downstream and subsequently the flame propagated upstream until stabilised around the bluff-bodies.
Since the ignition process (known as \textit{light-around}) involves an azimuthal motion of the flame, which gives rise to azimuthal pressure waves, the ignition transients may have an impact on the start-up of azimuthal instabilities by acting as \textit{an initial transverse forcing}. 

\begin{figure*}[!h]
\small
\centering\ifx\mycmd\undefined
\includegraphics[width=\textwidth]{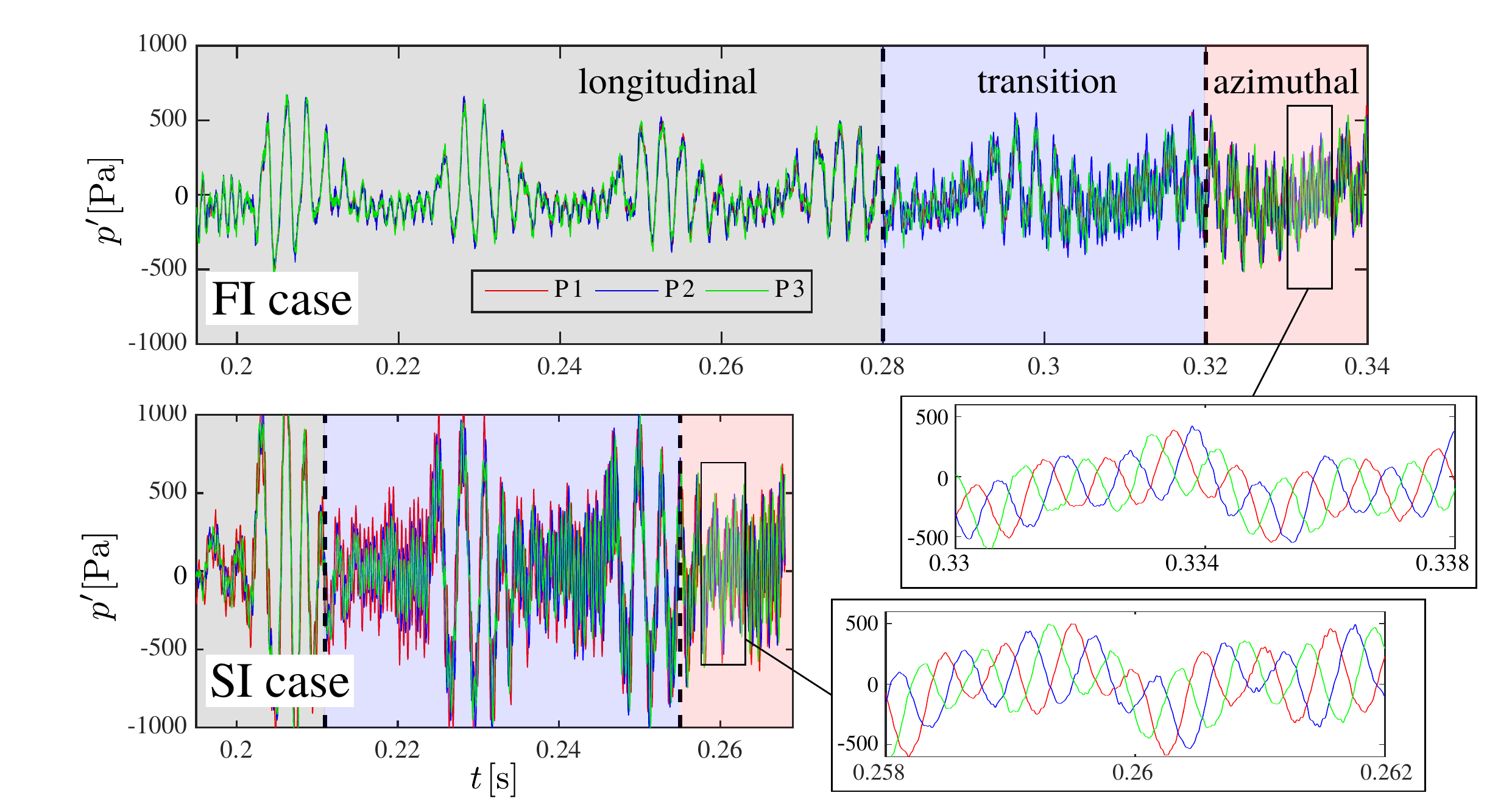}
\fi
\caption{Pressure time series for two different ignition cases (for adiabatic simulations). Probes P1, P2 and P3 are marked in Fig.~\ref{fig:exp_setup} (upper tapping position).  } 
\label{fig:pPr}
\end{figure*}

To shed light on this, two separate LES cases, referred to as full ignition (FI) and spark ignition (SI) respectively hereafter, are performed based on the same initial fields. 
Figure~\ref{fig:pPr} presents the computed time series of the pressure fluctuation at probes P1, P2 and P3, which are located in the injector tubes, 120$^\circ$ apart in the azimuthal direction and 45~mm upstream of the dump plane.
For the FI case, the entire annular chamber is ignited by duplicating the single-sector solutions. 
The ignition location for SI is on the centreline of one of the bluff-bodies and the height is the same as the inner wall (see marked in Fig.~\ref{fig:exp_setup}). 
For both the FI and SI cases, a transient evolution from ignition to fully excited azimuthal mode is observed in the LES. Shortly after numerical ignition at $t=0.195$~s, significant longitudinal oscillations ($>$500~Pa) appear and the three pressure signals collapse onto one another. This period is quite long for the FI case and the large oscillations occur several times spanning over almost 0.1~s. For the SI case, the longitudinal oscillations reach a much higher amplitude, about twice as that for the FI case, which consequently initiate the transition period at a much earlier time about 0.015~s after ignition. 
During the transition period, mixed longitudinal and azimuthal oscillations co-exist and the length of this period is quite similar between the two ignition cases, although the amplitudes are generally higher for the SI case. 
Finally, dominant azimuthal modes are established in both cases showing very similar frequency and amplitude as can be seen in the zoomed signals in Fig.~\ref{fig:pPr}. 
This ignition sensitivity test suggests that although azimuthal instability can be reproduced using either ignition methods, the onset of the azimuthal modes is strongly influenced by the transient ignition processes and the spark ignition approach followed by a natural light-around process leads to a significantly faster establishment of the instability. This has a significant implication for LES of azimuthal instabilities as the computational cost can be substantially reduced. 

\subsection{Comparison with the experiments} \label{subsec:comparison}

The statistics obtained from experiments and simulations for the spinning azimuthal modes are compared in this subsection. It should be noted that the start-up process, i.e., from ignition until limit-cycle oscillation, was observed to be quite long (up to almost 10~s), although the azimuthal instability already appears in the first several hundreds of milliseconds. During this process, the oscillation frequency remains nearly the same from the beginning, whereas the amplitude ramps up gradually. 
As simulating the entire process for 10~s is computationally impractical, the same sampling period, 400 to 500~ms after ignition, was chosen for both the experimental and LES pressure data to make an appropriate comparison. 

\begin{figure*}[!h]
\small
\centering\ifx\mycmd\undefined
\includegraphics[width=0.5\columnwidth]{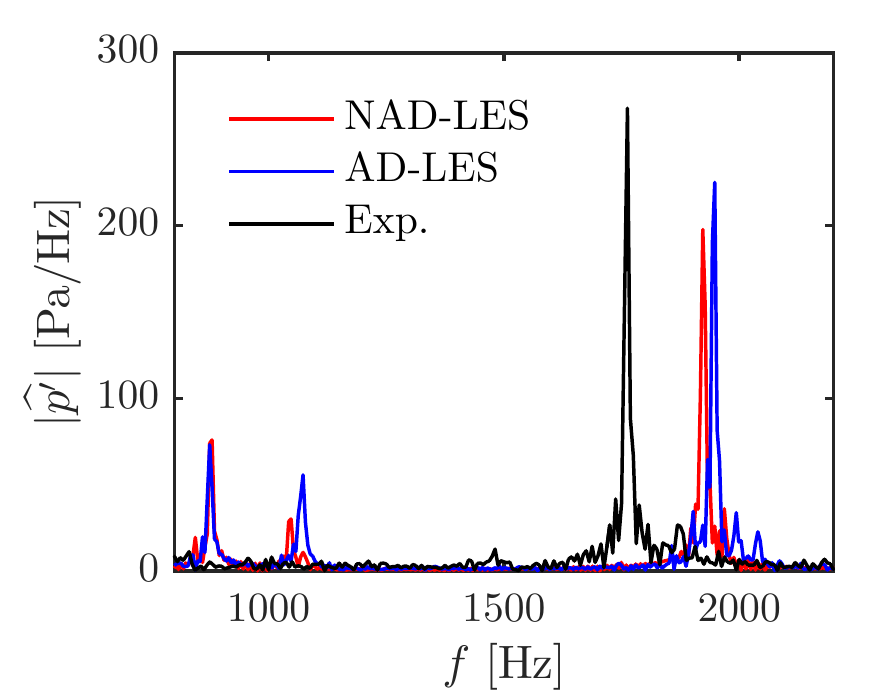}
\hspace{5mm}
\includegraphics[width=0.45\columnwidth]{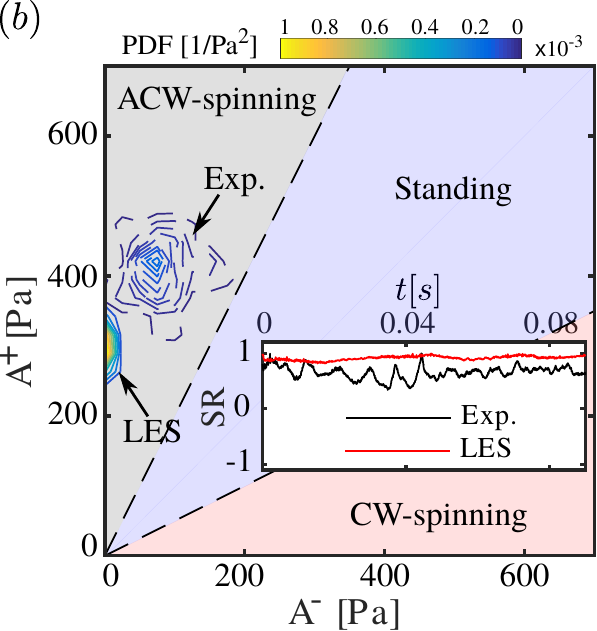}
\fi
\caption{Comparison of measured and computed (a) pressure power spectra and (b) joint PDF of the azimuthal wave amplitudes with the spin ratio (SR) plotted in the inset.  } 
\label{fig:p_PSD_PCI21}
\end{figure*}

Figure~\ref{fig:p_PSD_PCI21}a compares the measured pressure power spectra and those obtained from LES with the AD and NAD approaches described earlier in Section~\ref{sec:modelling}. 
It can be seen that both models are able to capture the azimuthal instability although the predicted frequencies are 100$-$150~Hz higher than the measured value at around 1800~Hz. The reason for this difference is that the acoustic mode structure dictating the excited mode lies in the upstream plenum, having one wavelength in both the longitudinal and azimuthal directions. Therefore, omitting the conical volume (mentioned earlier in Fig.~\ref{fig:exp_setup}) effectively resulted in a shorter longitudinal length and hence a higher fundamental frequency. More details on this will be discussed later in Section~\ref{sec:low_order}.

The amplitude in Fig.~\ref{fig:p_PSD_PCI21} is also under-predicted by both LES models by 5$-$8\%.
By including the heat loss effects, the NAD model gives a slightly lower frequency and smaller amplitude, but the overall difference is marginal compared to the adiabatic approach. 
This suggests that unlike the compact swirling flames~\cite{BauerheimPCI15}, the bluff-body stabilised flames are much longer and their response to acoustic perturbations are less affected flame root dynamics induced by the wall heat losses. 
This is also consistent with previous numerical studies on the FTF/FDF of a single bluff-body flame~\cite{BalachandranAKDM05}, see e.g. Refs.~\cite{HanCNF15,RuanCST16}, where the non-linear flame response was well captured by adiabatic models. 
As the two LES models give similar results, further analyses hereafter are performed only for the adiabatic case for simplicity. 

\begin{figure*}[!th]
\small
\centering\ifx\mycmd\undefined
\includegraphics[width=\textwidth]{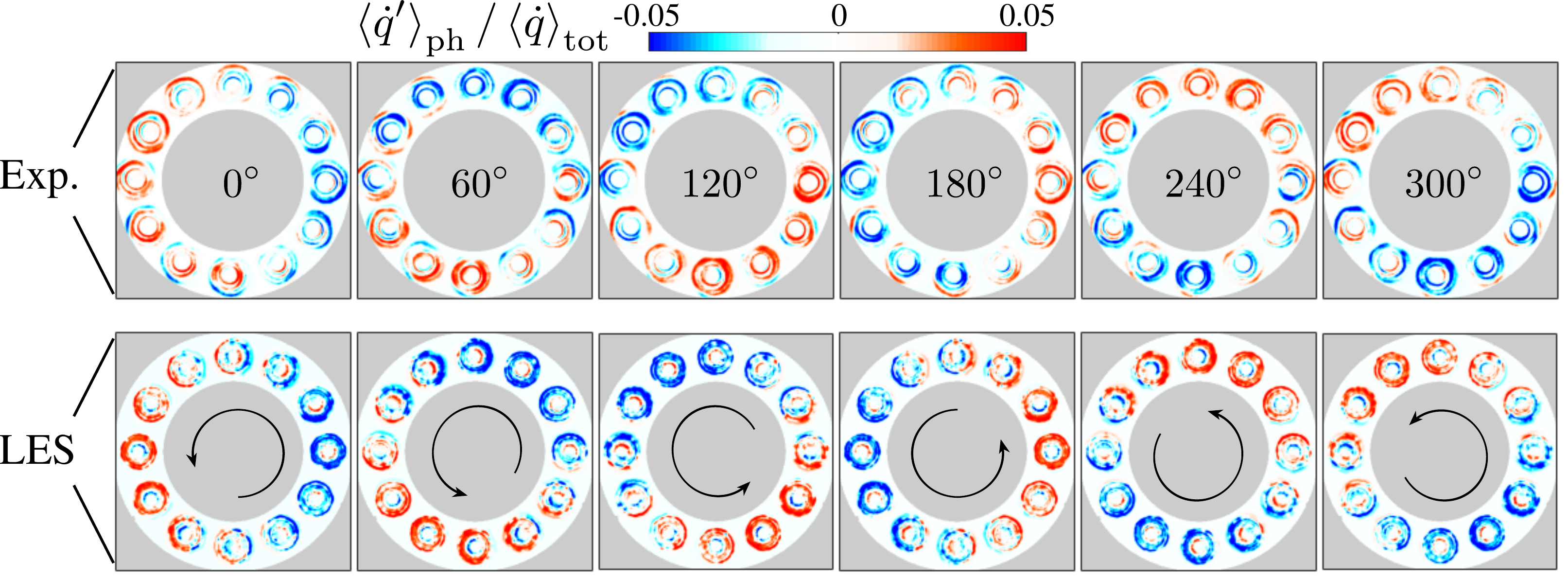}
\fi
\caption{Comparison of phase-averaged fluctuating heat release rate (FHRR) obtained from overhead OH$^*$ chemiluminescence measurements (top) and line-of-sight integration of LES results (bottom).  } 
\label{fig:phaseAvg}
\end{figure*} 

To compare the azimuthal mode characteristics, the measured and computed pressure data at probes P1, P2 and P3 (see Fig.~\ref{fig:exp_setup}) are processed to obtain the super-position of azimuthal wave amplitudes following the least-square fitting procedure used in Refs.~\cite{MazurPCI19,WorthCNF13}. The amplitudes of anti-clockwise and clock-wise spinning waves are denoted as $A^+$ and $A^-$ respectively. 
Based on these amplitudes, the spinning and standing modes can be well defined and the joint PDFs of $A^+$ and $A^-$ obtained from the measurement and LES are compared in Fig.~\ref{fig:p_PSD_PCI21}b. A good agreement can be observed with both exhibiting an evident ACW-spinning mode. While the experimental PDF shows a considerable scatter, the LES results fall very close to the vertical axis with high PDF values, suggesting a stronger preference in the ACW spinning direction. This may be indicative of less noise in the LES~\cite{GhirardoPRSA13,GhirardoARXIV19}.  It is also to be noted that in this symmetric burner configuration (without bulk swirl), there should be equal probabilities for CW or ACW modes. Some level of asymmetry due to manufacturing or other uncertainties resulted in a preferences for ACW observed over many experimental runs. However, the geometry is perfectly symmetric in the LES, which should prefer neither directions, and the ACW mode observed is likely to be a coincidence for a single LES run. Using the definition given in~\cite{BourgouinASME13}, the spin ratio can be calculated as: ${\rm SR} = (A^+-A^-)/(A^+ + A^-)$. The measured and computed SR are plotted in the inset of Fig.~\ref{fig:p_PSD_PCI21}b. 
Consistently with the joint PDFs, both the LES and experiment give SR values close to 1 throughout the sampling period, featuring a reasonably \textit{stable} ACW spinning mode. Mode switching is not observed for this case due to the relatively high equivalence ratio and bulk velocity, which was shown to prefer spinning modes~\cite{MazurPCI19}.

The strength and distribution of the fluctuating heat release rate (FHRR) and its response to the spinning azimuthal waves are of prime importance for understanding azimuthal instabilities. 
Thus, phase-averaged normalised FHRRs for six angles ($0^\circ$, $60^\circ$, ..., $300^\circ$) obtained from the experimental OH$^*$ images and LES results are compared in Fig.~\ref{fig:phaseAvg}.
The maximum heat release fluctuation in the LES is about 7\%, which agrees well with the experimental value of 6\%.
The previously reported~\cite{MazurPCI19} crescent shaped structures around both sides of the flame is also observed here in both the measurements and simulations. 
Despite the considerable noise in the experimental data, a quite symmetric FHRR distribution is seen with half of the burners forming a positive semicircle and the other half forming a negative one. This structure spins as one moves along the cycle from left to right in the figure. 

The above comparisons for the pressure and heat release rate show an overall good agreement between the experiment and simulation, and this suggests that the spinning mode and its major characteristics are well captured in the LES.

\subsection{DMD analysis of the azimuthal mode} 

Dynamic mode decomposition (DMD)~\cite{SchmidJFM10} is a powerful technique to study coherent spatial-temporal structures in dynamic systems. Compared to proper orthogonal decomposition (POD), DMD generates a set of modes associated with their own oscillation frequency, amplitude and growth/decay rates, which are desired for a better interpretation of combustion dynamics~\cite{HuangAIAA16}. Hence, DMD is applied here to further investigate the spinning mode azimuthal instability.
 
The DMD algorithm implementation of~\cite{SchmidJFM10} is followed to process 516 snapshots collected from continuous measurements and simulations spanning over about 30~ms. This corresponds to approximately 50 cycles of the azimuthal mode. 
Note that the experimental snapshots are 2D line-of-sight OH$^*$ images taken from the top of the burner, and the LES snapshots are 3D heat release rate field for the entire annular combustion chamber. Prior to the DMD computation, the LES results are integrated in the streamwise direction and then projected to the same 2D plane as the OH$^*$ images. 
The DMD mode growth rates for the peak frequency (about 1800 and 1900~Hz for the experiment and LES respectively) are near-zero negative values, suggesting that limit-cycle oscillations have been established before the sampling period. It is worth remarking here that in the DMD spectrum for the LES, two eigen-frequencies at about 900 and 1100~Hz also showed considerable amplitudes, which are consistent with the two small peaks seen in Fig.~\ref{fig:p_PSD_PCI21} and the low frequency content seen on the time series in Fig.~\ref{fig:pPr}. These are associated with the longitudinal modes of the combustor geometry. However, both of these DMD modes have large negative growth rates suggesting that they will eventually be damped out in a longer simulation. Nevertheless, the azimuthal mode of interest is well established already in the present simulation duration and hence is discussed next.  

\begin{figure}[!h]
\small
\centering\ifx\mycmd\undefined
\includegraphics[width=0.7\columnwidth]{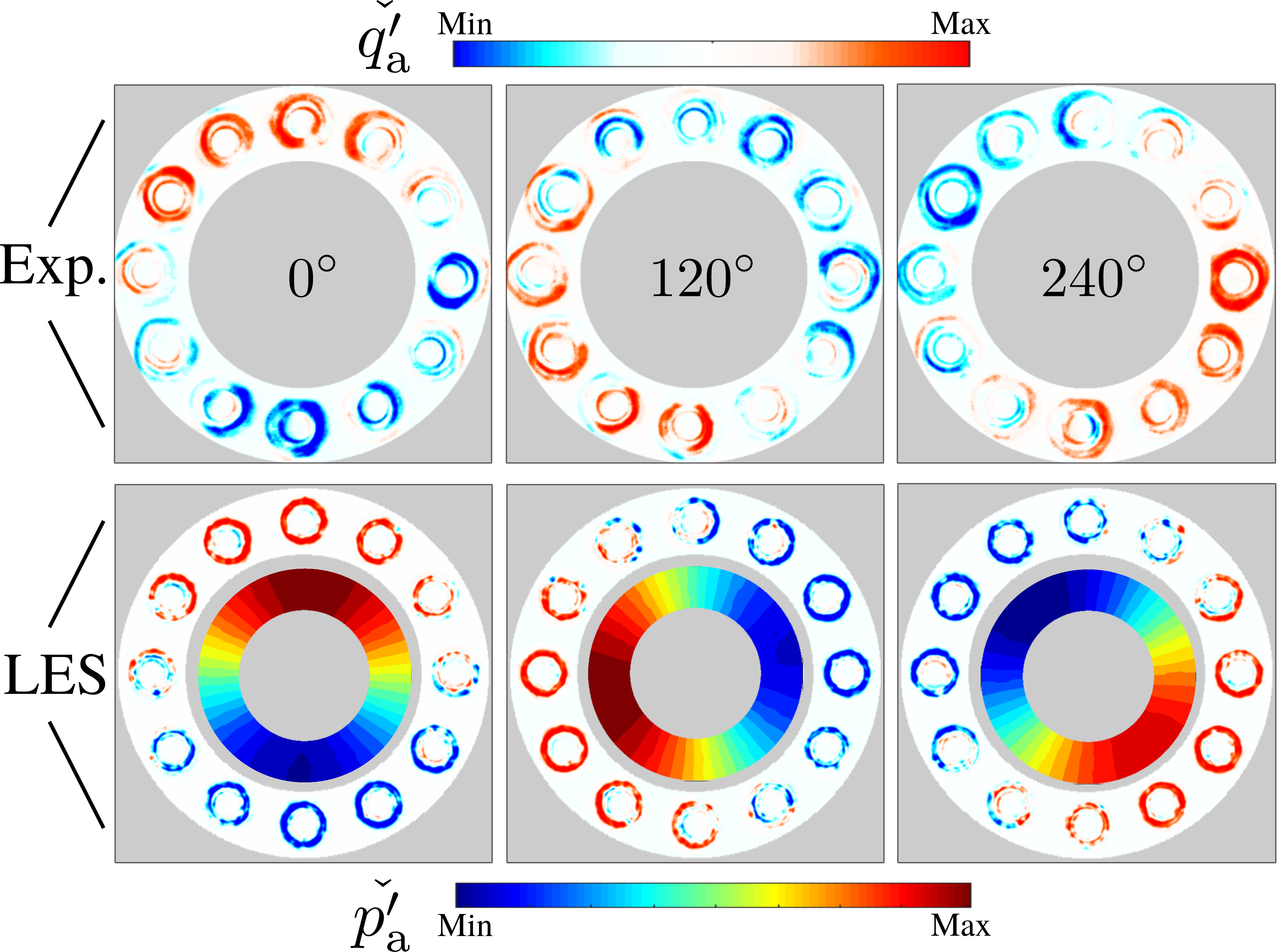}
\fi
\caption{Phase-shifted DMD reconstruction of the FHRR at the azimuthal mode frequency for three phase angles. The corresponding combustion chamber pressure mode is shown for the LES. } 
\label{fig:DMD_q}
\end{figure} 

\begin{figure}[!h]
\small
\centering\ifx\mycmd\undefined
\includegraphics[width=0.7\columnwidth]{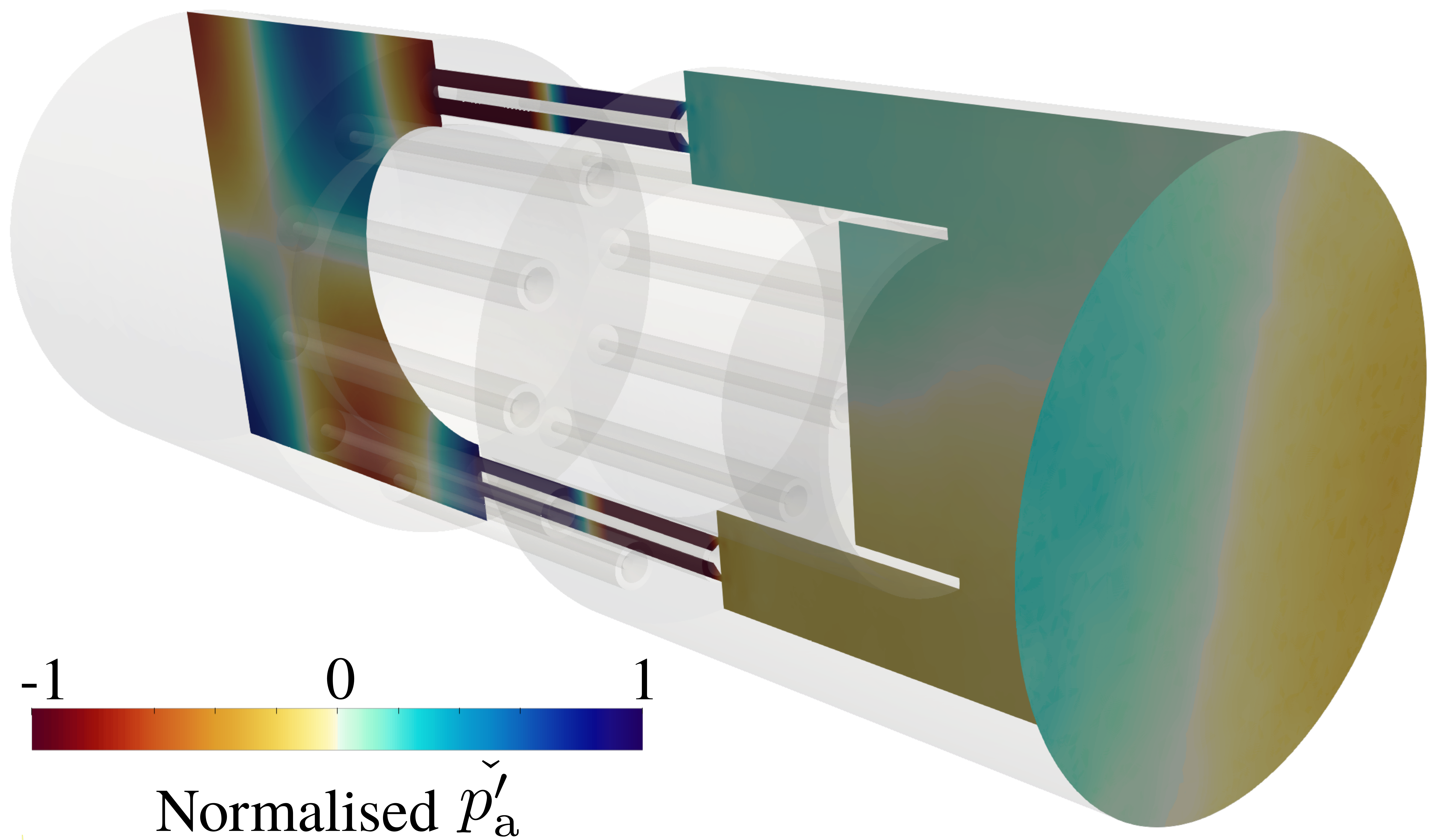}
\fi
\caption{DMD mode structure of the acoustic pressure reconstructed at the azimuthal mode frequency from the LES.   } 
\label{fig:DMD_p_3D}
\end{figure}

The DMD reconstructions of the heat release rate at the azimuthal oscillation frequencies are shown in Fig.~\ref{fig:DMD_q} for three phase angles. The reconstructed pressure field (not available from the experiments) is also shown for the LES. Note that these phases here do not necessarily correspond to those shown in Fig.~\ref{fig:phaseAvg}, because there is an arbitrary phase shift during the DMD process. 
It is seen that the reconstructed modes show relatively less noise compared to the simple phase averages, particularly for the LES. Again, good agreement is obtained here for the DMD modes for FHRR, which further confirms that the spinning mode behaviour is well captured in the LES. 
It is also seen that the DMD modes for pressure and FHRR seem to be synchronised in the annular chamber. Finally, the DMD mode structure for the LES domain is shown in Fig.~\ref{fig:DMD_p_3D}. 
The mode structure features a first longitudinal/first azimuthal mixed mode in the upstream plenum, a first longitudinal mode with the pressure node at the middle of the premixer, and a quarter-wave longitudinal/first azimuthal mixed mode in the combustion chamber. These are found to be quite similar to that computed in~\cite{BauerheimPCI15} using a Helmholtz solver. However, evident differences are observed here for the downstream combustion chamber outlet (top of the outer wall), where the LES acoustic pressure retains a considerable value compared to the $p^{\prime}=0$ condition imposed in the Helmholtz solver.
This downstream mode structure may be essential for self-excited azimuthal modes to occur.
It also implies that downstream acoustic boundary condition could have a major role on the overall stability and azimuthal modes, and hence should be handled appropriately in the analytical and low-order modelling approaches. 

\section{Low-order modelling}\label{sec:low_order}

\begin{figure*}[!th]
\begin{center}
\includegraphics[width=\textwidth]{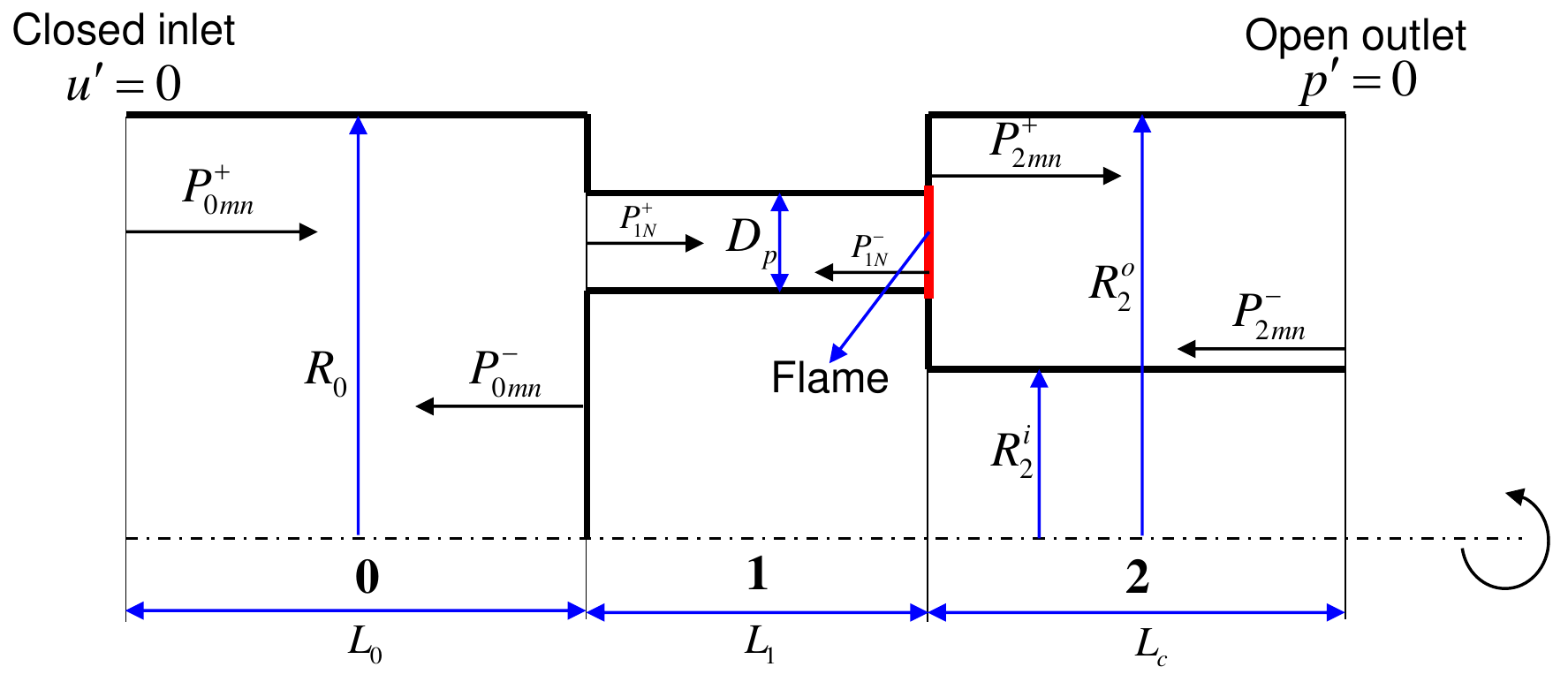}
\end{center}
\caption {Sketch of the half cross-section view of the simplified combustor. The sketch is not to scale.}\label{Sketch_premixed_combustor}
\end{figure*}

In order to further understand the azimuthal mode predicted in the LES and its dependency on the upstream plenum geometry, a low-order modelling analysis is performed. The model combustor geometry is simplified to establish an analytical network acoustic model, as shown in Fig. \ref{Sketch_premixed_combustor}. 
Note that the network model considers the system acoustics in the three-dimensional space (see Ref.~\cite{ZhangAIAA20} for details) and the sketch here only shows a half cross-sectional view of the 3D model.  
We consider the pressure disturbance in the form of 
\begin{equation}
	p'=\mathrm{Re}[\hat{p}\,\mathrm{e}^{\mathrm{i}\omega t}],
\end{equation}
where the operator $\mathrm{Re}[\cdot]$ takes real part of the complex value. 
The premixers are so thin that the waves inside can be assumed to be planar waves, written as
\begin{equation}
\hat{p}_{1N}(x)=P_{1N}^{+} \mathrm{e}^{\frac{-\mathrm{i} \omega x}{\overline{c}_1+\overline{u}_{1N}}}+P_{1N}^{-} \mathrm{e}^{\frac{\mathrm{i}\omega (x-L_1)}{\overline{c}_1-\overline{u}_{1N}}},
\label{p_premixer}
\end{equation}
\begin{equation}
\hat{u}_{1N}(x)=\frac{1}{\overline{\rho}_1  \overline{c}_1}[P_{1N}^{+} \mathrm{e}^{\frac{-\mathrm{i}\omega x}{\overline{c}_1+\overline{u}_{1N}}}-P_{1N}^{-} \mathrm{e}^{\frac{\mathrm{i}\omega (x-L_1)}{\overline{c}_1-\overline{u}_{1N}}}],
\label{u_premixer}
\end{equation}
where $P_{1N}^{+}$ and $P_{1N}^{-}$ denote the complex amplitudes of the downstream and upstream propagating waves respectively, and $N$ denotes the sequence number of the premixers. The axial coordinate $x$ is relative to the left end of each network element (0 for plenum, 1 for premixer and 2 for combustion chamber). The local speed of sound and mean velocity are denoted by $\overline c$ and $\overline u$ respectively, with the subscript corresponding to the given network element. The waves in the plenum and combustion chamber are three-dimensional, and can be obtained from the wave equation using common notations: 
\begin{flalign}
\hat{p}_j(\mathbf{r})=  \sum_{m=-\infty}^{m=+\infty}\sum_{n=1}^{+\infty} & \psi_{jm}(z_{jmn} r) \times \\  \nonumber
& [P_{jmn}^{+} \mathrm{e}^{i k_{jmn}^{+}x} +P_{jmn}^{-} \mathrm{e}^{i k_{jmn}^{-} (x-L_{j})} ], 
\label{pc}
\end{flalign}
\begin{flalign}
\hat{u}_j(\mathbf{r})=-\frac{1}{\overline{\rho}_j}\sum_{n=1}^{+\infty}\psi_{jm}(z_{jmn}r) &[\frac{k_{jmn}^{+} P_{jmn}^{+} \mathrm{e}^{\mathrm{i} k_{jmn}^{+}x}}{\omega+k_{jmn}^{+} \overline{u}_j}  \\ \nonumber 
&+\frac{k_{jmn}^{-} P_{jmn}^{-} \mathrm{e}^{\mathrm{i} k_{jmn}^{-}(x-L_{j})}}{\omega+k_{jmn}^{-} \overline{u}_j} ],
\label{uc}
\end{flalign}
where $j=0,2$ and $\psi$ denotes the eigenfunction of the circular duct ($j=0$) or annular duct ($j=2$), $m$ the azimuthal mode number and $n$ the radial mode number. The wavenumbers $k_{jmn}^{\pm}$ can be obtained by dispersion equations \cite{ZhangAIAA20}.

Since the combustion zone is nearly isobaric, the flame can be seen as compact and is located at the dump plane~\cite{DowlingS03}. The convective terms involving vorticity and entropy are neglected. Hence, the energy conservation across the flame is reduced to 
\begin{equation}
\frac{\gamma}{\gamma-1} \frac{\partial}{\partial x_j}(\overline{p}\hat{u}_j)=\hat{q}.
\label{energy_equation}
\end{equation}
The unsteady heat release rate term $\hat{q}$ using the linear $n-\tau$ flame transfer function (FTF) is obtained from the LES (full-annular adiabatic case), with the form
\begin{equation}
\frac{\hat{q}_N}{\overline{q}_N}=k_f\frac{\hat{u}_{1N}}{\overline{u}_{1N}}e^{-\mathrm{i} \omega \tau},\;x=L_0+L_1,
\label{unsteady_heat_release_per_premixer}
\end{equation}
where the gain $k_f=1.1$ and the phase difference $\omega_\tau=1.77$. The reference point for the velocity fluctuations was taken to be 60~mm upstream of the dump plane to keep clear of the inlet turbulence disturbances. It is worth remarking that the choice of this location has a quite considerable effect on the above FTF values due the standing mode inside the premixers. Also note that the FTF is only reliable at the self-excited frequency of the LES results. Equation (\ref{energy_equation}) is integrated over the ($r$, $\theta$) plane~\cite{AkamatsuASME01}. Then we have the closed equations that characterize the disturbances in the model combustor
\begin{equation}
\mathbf{X}(\omega)\mathbf{P}=\mathbf{0}.
\label{system_equation}
\end{equation}
By solving $\det[\mathbf{X}]=0$, we have the complex eigenfrequency $\omega=\omega_r+\mathrm{i}\omega_i$ of the system, where $\omega_r$ represents the natural frequency, and $-\omega_i$ represents the growth rate. The positive growth rate means the system is unstable, while the negative growth rate means stable.
Refer to Ref.~\cite{ZhangAIAA20} for elaborate modelling details.

\begin{table}[htbp]
\caption{\label{tab:table1} Model geometrical parameters.}
\centering
\begin{tabular}{lcccccc}
\hline
Parameters&  Symbols & Values\\\hline
Length of plenum & $L_0$ & 0.2 m \\
Radius of plenum & $R_0$ & 0.106 m\\
Length of premixer & $L_1$ & 0.155 m\\
Cross-section area of premixer & $A_p$ & 0.00015 $\mathrm{m}^2$\\
Length of combustion chamber & $L_c$ & 0.3 m\\
Outer radius of combustion chamber & $R_2^o$ & 0.106 m\\
Inner radius of combustion chamber & $R_2^i$ & 0.0635 m\\
 \hline
\end{tabular}
\label{geometrial_annular_pipe}
\end{table}

\begin{table}[htbp]
\caption{\label{tab:table1} Frequency variation against the plenum length.}
\centering
\begin{tabular}{lcccccccc}
\hline
$L_0/\mathrm{m}$&  0.12 & 0.14 & 0.16 & 0.18 & 0.20 & 0.22 & 0.24\\ \hline
$f/\mathrm{Hz}$&  3009 & 2637 & 2386 & 2115 & 1900 & 1817 & 1712 \\
$\mathrm{Growth rate}/\mathrm{(rad/s)}$& $\times$ & $\times$ & $\times$ & $\times$ & 409.2 & $\times$ & $\times$ \\
 \hline
\end{tabular}
\label{freq_plenum_length}
\end{table}

The list of model parameters used in the above low-order calculation are given in Table~\ref{geometrial_annular_pipe}. 
The length of the plenum is varied while keeping all other parameter unchanged to investigate its effect on the instability mode characteristics. 
Note that the length used for the LES mesh is 0.2~m (see Fig.~\ref{fig:exp_setup}). 
The modelling result for the frequency and growth rate of the corresponding mode versus the plenum length is presented in Table~\ref{freq_plenum_length}. 
The frequency of 1900 Hz for the plenum length equals 0.2 m agrees very well with LES result of 1910 Hz. 
Since the FTF is only reliable at the excited mode frequency, only the growth rate at 1900 Hz is presented. The predicted growth rate is positive implying that the system is unstable, which is again consistent with the LES. 
It can be also seen that the predicted frequency monotonically decreases with the increasing plenum length, showing a strong longitudinal dependency.
This provides an explanation for the over-predicted instability frequency in the LES because an effectively shorter length is considered without including the bottom conical inlet volume. 
The height of this cone is about 60~mm, which would give an effective length of 20~mm if converted to a cylindrical volume with the same circular area. 
Consequently, the corresponding plenum length $L_0=0.22$~m gives a frequency of 1817~Hz, which is close to the measured value around 1800~Hz. 
This observation suggests that one should be extremely careful while simplifying the burner geometry for either LES or low-order modelling as small details may have significant impact on the system instability behaviour. 

\begin{figure}[htbp]
\begin{center}
\includegraphics[width=0.7\columnwidth]{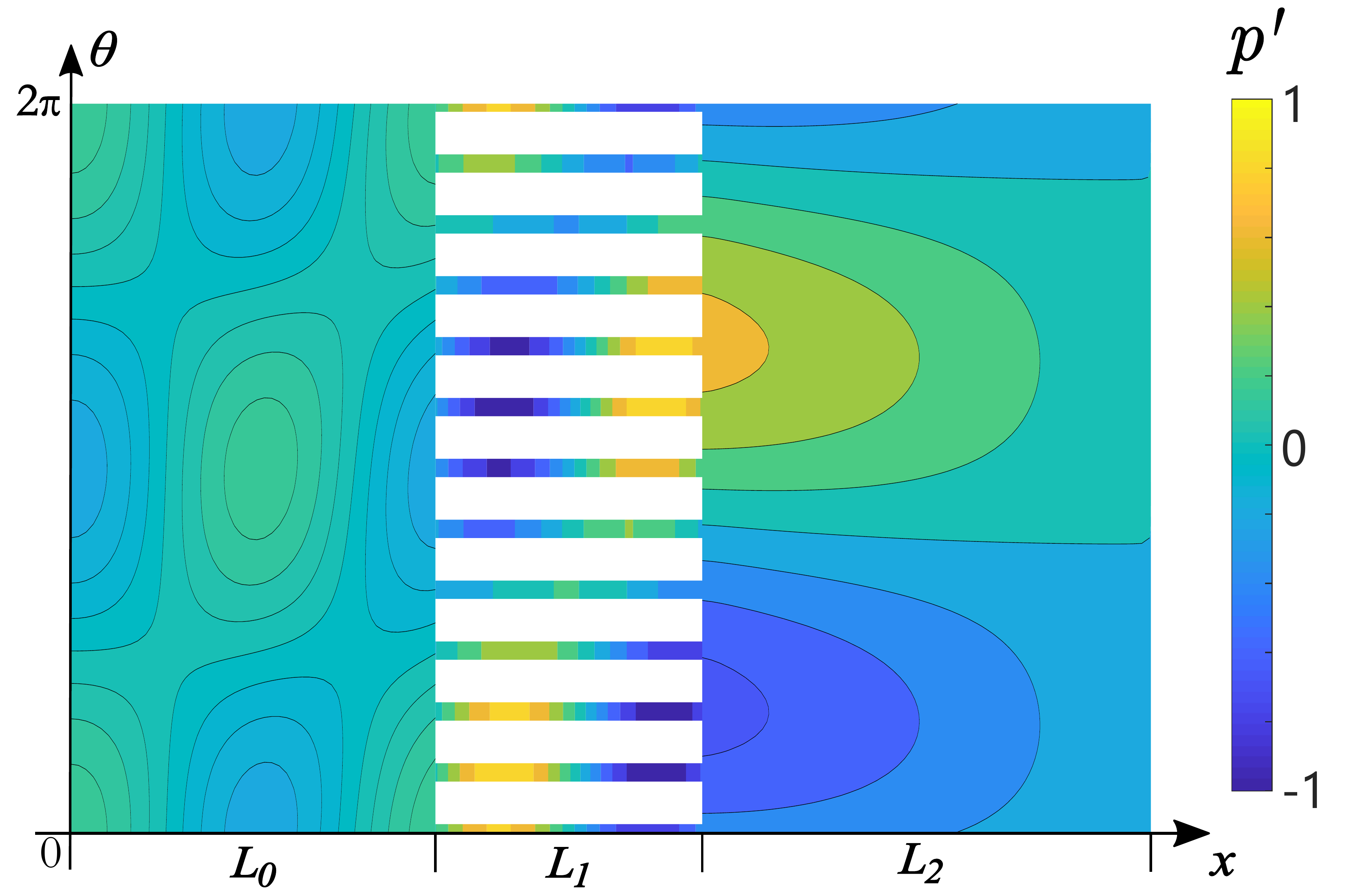}
\end{center}
\caption {Pressure mode shape at the mean radius of the model combustor when $L_0=0.2$ m.}\label{Mode_shape}
\end{figure}

\begin{figure}[htbp]
\begin{center}
\includegraphics[width=0.6\columnwidth]{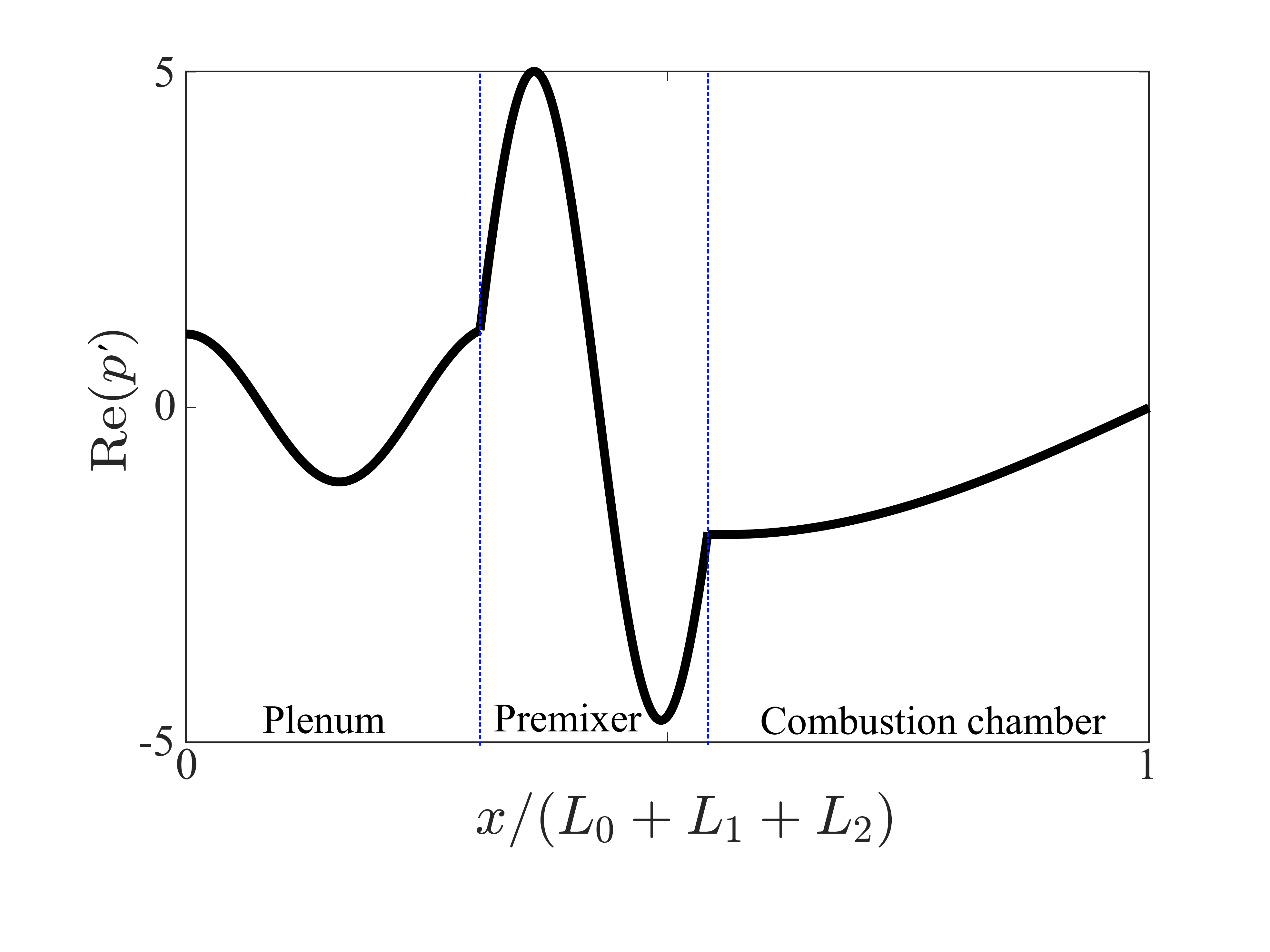}
\end{center}
\caption {Axial pressure mode shape at the mean radius of the model combustor when $L_0=0.2$ m and $\theta=0^{\circ}$.}\label{Mode_shape_axial}
\end{figure} 

The predicted mode shape by the low-order model is shown in Fig.~\ref{Mode_shape}. The overall distribution also matches well with that obtained by the LES (see Fig.~\ref{fig:DMD_p_3D}). Figure~\ref{Mode_shape_axial} depicts the axial pressure mode shape at the mean radius of the model combustor when $L_0=0.2$ m and $\theta=0^{\circ}$. It is seen that one wavelength exists in both the plenum and premixer, while the pressure amplitude in the premixer is larger than that in the plenum. Although the combustion chamber occupies the largest length scale, the axial pressure variation there is the smallest due to the higher temperature and hence larger speed of sound and wavelength. 
A notable difference between the mode shapes predicted by LES and low-order model is that the LES gives a higher amplitude in the plenum compared to that in the combustion chamber, whereas it is the opposite in the low-order modelling. It is speculated that this is caused by the simplified outlet boundary geometry condition, i.e. a flat open outlet shown in Fig.~\ref{Sketch_premixed_combustor}, whereas in the present LES and previous Helmholtz simulation~\cite{BauerheimPCI15} the outer wall is much higher than the inner wall (see Fig.~\ref{fig:DMD_p_3D}). The former is fully reflective while the latter is not. As a result, the amplitude in the combustion chamber given by the low-order model is relatively higher than that obtained from LES and Helmholtz solver. Unfortunately, experimental measurement of the pressure amplitudes in the plenum and combustion chamber is unavailable for validation. Nevertheless, the LES and low-order modelling results presented here provide useful information for the pressure mode structures inside the combustor, which complements the experimental observations with useful physical insights. Future works will be directed towards understanding and predicting the various modal dynamics and explore the effects of the downstream boundary conditions.


\section{Conclusions}\label{sec:conclusions}

A comprehensive numerical study including LES and low-order modelling of the azimuthal mode thermoacoustic instability in an annular combustor with non-swirling flows is presented. 
Two LES flamelet models, with and without considering the heat losses, are used to investigate non-adiabatic effects on the azimuthal instability. The azimuthal modes are captured by both models and only marginal differences are observed for the frequency and amplitude. The ability of flamelet-based LES approach to capture self-excited azimuthal modes is demonstrated for the first time in this work. Although the amplitudes are predicted well, the computed frequencies are higher than the measured value (about 1800~Hz) by 8$-$12\%, which is due to the simplified upstream plenum geometry. The subsequent low-order modelling results show that when the effectively plenum length is matched the frequency agrees very well with the measurement. 

Analysis of the pressure data from experiments and LES indicates an anti-clockwise spinning mode with the spin ratio close to positive unity. Comparison of phase-averaged heat release rate fluctuations shows a good agreement for this spinning mode. DMD is applied to further investigate the modal behaviours at the azimuthal mode frequency. A non-zero combustor outlet mode structure is observed for the pressure in the LES, which suggests that the downstream geometry may play an important role in establishing the azimuthal modes in this combustor. The LES observation is further supported by the low-order modelling results showing very similar mode shapes in the plenum and premixers, but different in the combustion chamber due to a simplified outlet boundary. The higher amplitude in the combustion chamber indicates that a fully reflective open boundary condition may be inadequate and possible improvements will be explored in future works. 

\section*{Appendix A. Assessment of mesh sensitivity}
\renewcommand{\thefigure}{A.\arabic{figure}}

\begin{figure}[h!]
   	\centering\ifx\mycmd\undefined
	\includegraphics[width=0.95\columnwidth]{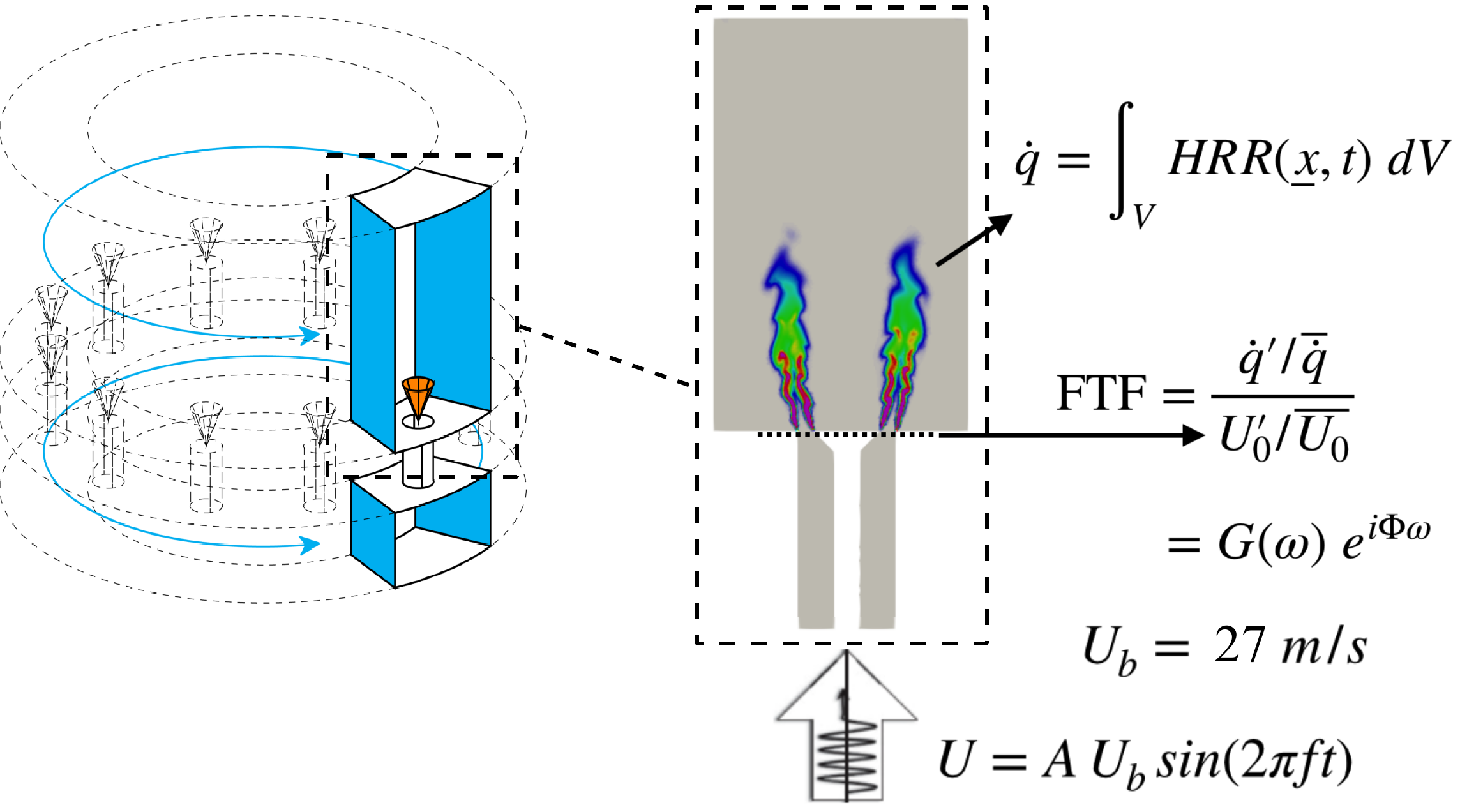}
	\fi\caption{Illustration of the single sector simulation with upstream velocity forcing. Left schematic adapted from \cite{MensahJEGTP16}, copyright 2016, with permissions from ASME.}
		\label{fig:schematic}
\end{figure}

To avoid the high computational cost of repeatedly running full annular combustor simulations, the mesh sensitivity is investigated using a single sector of the 12 burner configuration~\cite{MazurPCI19}. A schematic of the simulation setup is shown in Fig.~\ref{fig:schematic}.
The bulk velocity at the bluff-body exit is $U_b = 27$~m/s and the equivalence ratio is 0.9 for the premixed ethylene/air mixture at 300~K and 1~atm. Two meshes having 1.2 and 2.1 million tetrahedral cells are tested here and the typical grid spacing near the shear layer is about 0.6 and 0.3~mm respectively. 

\begin{figure*}[th!]
   	\centering\ifx\mycmd\undefined
	\includegraphics[width=\textwidth]{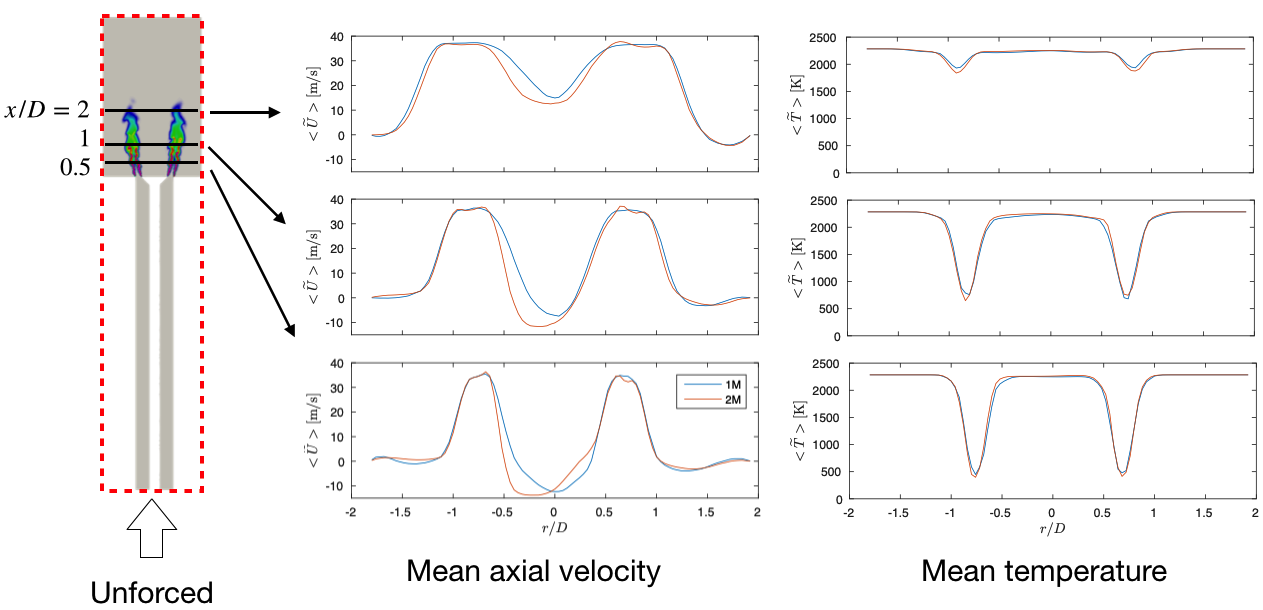}
	\fi\caption{Mean axial velocity and temperature comparisons for the unforced case. }
		\label{fig:meanFields}
\end{figure*}

First, the unforced reacting flow is simulated for the 1.2M and 2.1M meshes. After the flame is fully established, statistics are collected for about 4 flow-through-time, which is calculated based on the combustion chamber length of 80~mm and the bulk velocity. Figure~\ref{fig:meanFields} compares the computed mean axial velocity and temperature radial profiles at axial distances of $x/D=0.5$, 1 and 2. The bluff-body diameter is $D=13$~mm. As can be seen, the two simulations give very similar results for the mean axial velocity field at all three streamwise positions except for some considerable differences in the bottom of the recirculation zone. This is probably due to the relatively short statistics sampling time for the 2.1M case as indicated by the asymmetry. The difference in the temperature fields is almost negligible suggesting a good grid convergence. 

\begin{figure}[h!]
   	\centering\ifx\mycmd\undefined
	\includegraphics[width=\columnwidth]{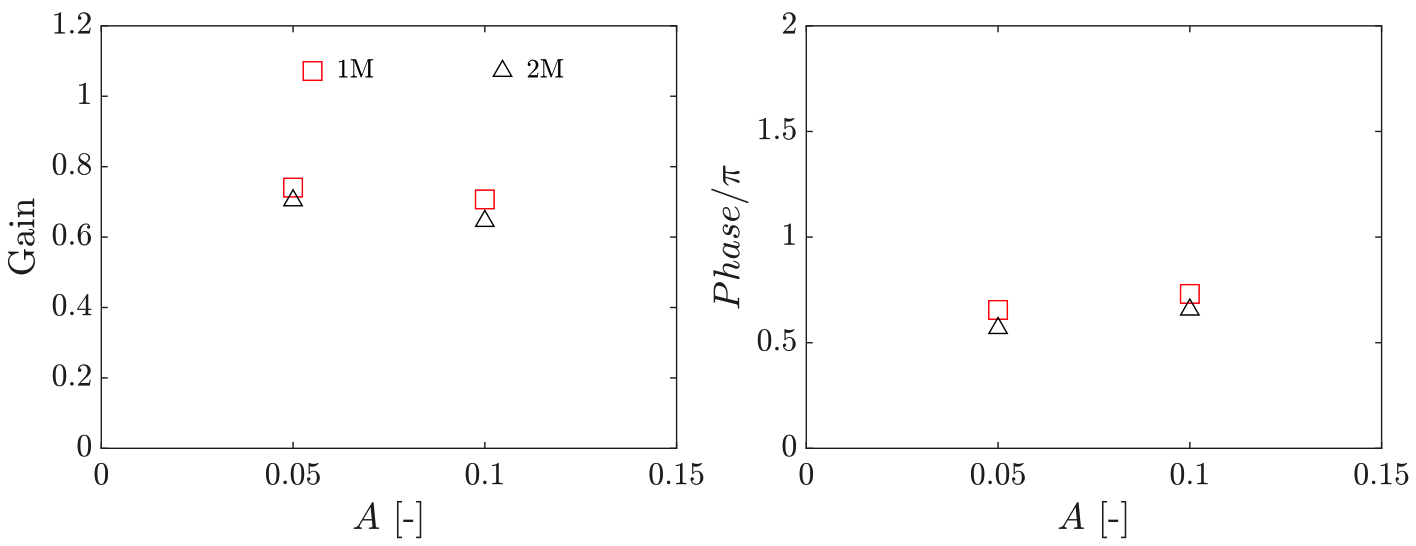}
	\fi\caption{Time series of the fluctuating velocity at the bluff-body exit (black) and the heat release response (red) to the excitation for the 1M (solid) and 2M (dashed) meshes.}
		\label{fig:FTF}
\end{figure}

Further grid sensitivity is assessed by performing LES with forced inlet velocity (See Fig.~\ref{fig:schematic}) to compare the flame transfer functions given by the two meshes. As the azimuthal modes are of  interests here, a fixed forcing frequency of 1800~Hz corresponding to the first azimuthal mode based on the perimeter of annular chamber is considered. Also, as the amplitude of pressure oscillation is below 1\% of the ambient pressure~\cite{BauerheimPCI15}, the two relatively low forcing amplitudes, 5\% and 10\% of the mean flow velocity respectively, are used for the forcing. 
\begin{figure}[h!]
   	\centering\ifx\mycmd\undefined
	\includegraphics[width=\columnwidth]{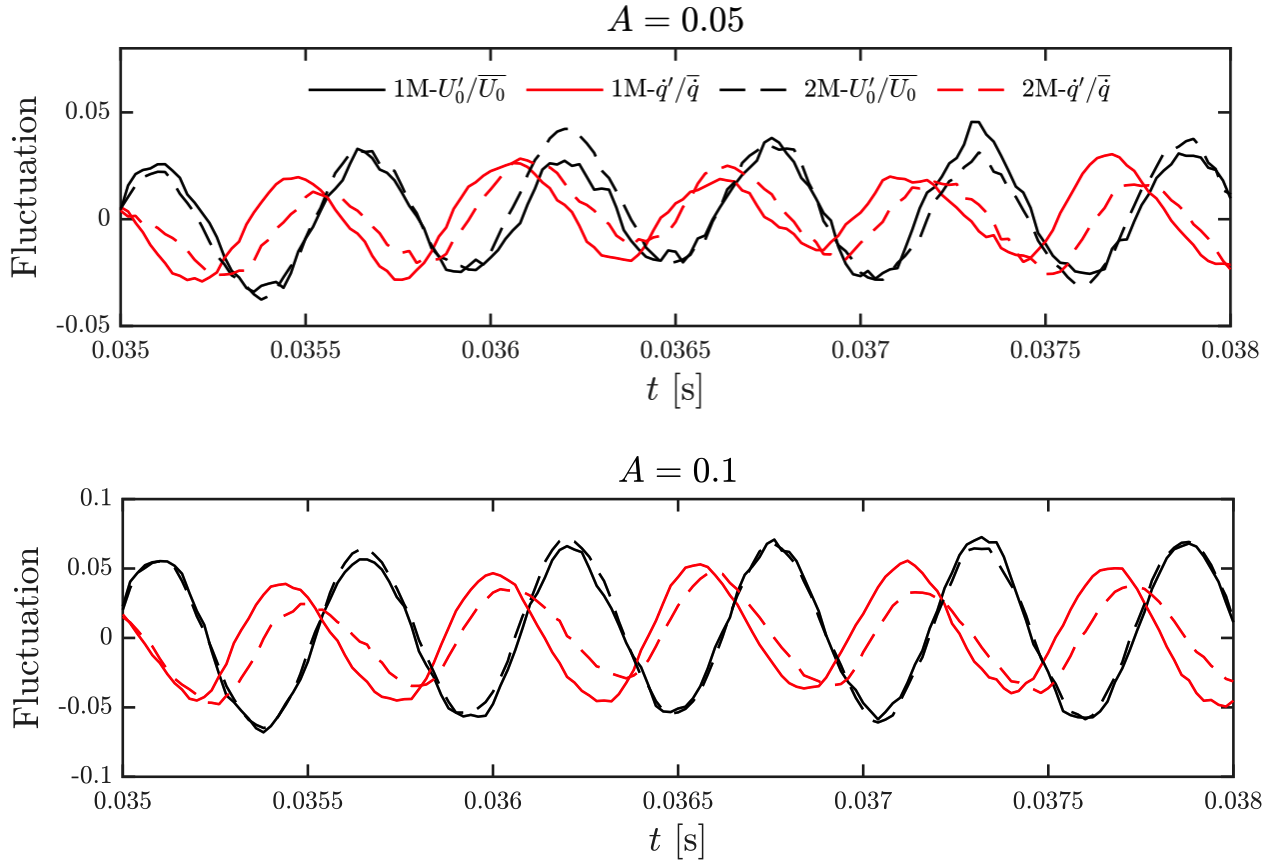}
	\fi\caption{Time series of the fluctuating velocity at the bluff-body exit (black) and the heat release response (red) to the excitation for the 1M (solid) and 2M (dashed) meshes.}
		\label{fig:forcedSignals}
\end{figure}
Figure~\ref{fig:forcedSignals} presents the time series (after transition) of the normalised fluctuating velocity at the bluff-body exit and heat release response to the excitation for the two meshes (see Fig.~\ref{fig:schematic} for the formulae used).
The overall fluctuating behaviour of the two signals are quite similar despite some slight differences in the heat release amplitude and the time delay (within 10\%). To show this in clearer manner, the gain and phase of the transfer functions are computed as shown in Fig.~\ref{fig:FTF}. Again, similar values are obtained from the two meshes for both the gain and phase. From 5\% to 10\% forcing amplitude, the gain slightly decreases and the phase increases, and this trend is also captured in both the 1.2M and 2.1M cases.

The above grid sensitivity study on the single sector suggests that the smaller 1.2M mesh is sufficient to capture the reactive flow fields and also the flame response to external forcing excitations. Thus, this grid distribution is applied to all 12 sectors for the annular combustion LES shown in the main text of this paper. 

\section*{Acknowledgements}
\label{Acknowledgments}
ZXC and NS acknowledge the support of Mitsubishi Heavy Industries, Takasago, Japan. This work used the ARCHER UK National Supercomputing Service (http://www.archer.ac.uk) with the computational time provided by the UK Consortium for Turbulent Reacting Flow. 

\bibliography{references_Zhi}

\end{document}